\documentclass[10pt,twocolumn,article]{IEEEtran}
\usepackage{times}
\usepackage{amsbsy}
\usepackage{latexsym}
\usepackage{amssymb}
\usepackage{mathtools,xparse}
\usepackage{amsmath,amssymb,amsfonts,graphicx,epsfig,amsthm}
\usepackage{times,amsbsy,latexsym,bm,color}
\usepackage[ansinew]{inputenc}
\usepackage{algpseudocode, algorithm}
\usepackage{setspace}
\usepackage{textcomp}
\usepackage{parskip}

\title{\huge Study of Coarse Quantization-Aware Block Diagonalization Algorithms for MIMO Systems with Low Resolution  \vspace{-0.025em}}
\author{Silvio F. B. Pinto $^1$ and Rodrigo C. de Lamare $^{1,2}$ \\
Center for Telecommunications Studies (CETUC) \\ $^1$ Pontifical
Catholic
University of Rio de Janeiro, RJ, Brazil.\\
$^2$ Department of Electronics, University of York, UK \\
Emails: silviof@cetuc.puc-rio.br, delamare@cetuc.puc-rio.br
\vspace{-0.25em}} 
\begin{document}
\maketitle
\begin{abstract}
It is known that the estimated energy consumption of digital-to analog converters (DACs) is around 30\% of the energy consumed by analog-to-digital converters (ADCs) keeping fixed  the sampling rate and bit resolution. Assuming that similarly to ADC, DAC dissipation doubles with every extra bit of resolution, a decrease in two resolution bits, for instance from 4 to 2 bits, represents a 75$\% $ lower dissipation. The current limitations in sum-rates of 1-bit quantization have motivated researchers to consider extra bits in resolution to obtain higher levels of sum-rates. Following this, we devise coarse quantization-aware precoding using few bits for the broadcast channel of multiple-antenna systems based on the Bussgang theorem. In particular, we consider block diagonalization algorithms, which have not been considered in the literature so far. The sum-rates achieved by the proposed Coarse Quantization-Aware Block Diagonalization (CQA-BD) and its regularized version (CQA-RBD) are superior to those previously reported in the literature. Simulations illustrate the performance of the proposed CQA-BD and CGA-RBD algorithms against existing approaches.

\end{abstract}


\section{Introduction}
\label{introduction}

Recent work in wireless communications has shown a great deal of progress in massive multiple-input multiple-output (MIMO) systems, which in case of transmissions using broadcast channels employ base-stations (BS) composed of a huge number of antennas. Nonetheless, the increasing numbers of antennas at the BS result in higher costs in terms of equipment and energy consumption. Thus, the design of effective and economical MIMO-based systems to provide coverage of geographical areas and cost-effective systems will require more energy-efficient and low-cost components \cite{Rusek,Larsson,Lu,Sarajlic,DeLamare_2013,Wence_cal}.

Despite the  progress  in  1-bit quantization \cite{Landau, Mezghani} with the aim of reducing  energy consumption in the large number of DACs used in massive MIMO, the achievable sum rates still remain low, which makes higher resolution quantizers with $b=2,3,4$ bits attractive for the design of precoders and receivers. Bussgang's theorem \cite{Bussgang} lets us express a Gaussian precoded signal that was quantized as a linear function of the quantized input and a distortion term which has no correlation with the input \cite{Sven1,Sven2,Sven3}. This approach makes possible the computation of sum-rates of Gaussian data \cite{Rowe}.

In this context, block diagonalization (BD) precoding methods and their variants \cite{Spencer1, Stankovic, Zu_CL, Zu, Sung,Wence} are known to be linear transmit approaches for multiuser MIMO (MU-MIMO) systems based on singular value decompositions (SVD), which provide excellent achievable sum-rates in the case of significant levels of multi-user interference. However, BD has not been considered with coarsely quantized signals so far.

Motivated by the relatively poor performance of  1-bit quantization of precoded signals applied to massive MU-MIMO systems, we propose coarse quantization-aware BD (CQA-BD) type precoders for signals quantized with an arbitrary number of bits in broadcast channels. Then, using Bussgang's theorem we derive expressions to compute the achievable sum-rates of the proposed CQA-BD type precoders. Simulations illustrate the excellent sum-rate performance of the proposed CQA-BD and CQA-RBD precoders against previously reported techniques.

This paper is structured as follows. Section \ref{sysmodel} briefly
describes the system model and
background for understanding the  proposed CQA-BD class algorithms. Section \ref{proposed_CQA_algorithms} presents the proposed CQA-BD type algorithms. In Section \ref{numerical_results}, we present and discuss numerical results whereas the conclusions are drawn in Section \ref{conclusions}.

\textit{Notation}: the superscript \textit{H} denotes the Hermitian transposition,  $\mathbb E[\cdot]$ expresses the expectation
operator, $\bm I_M$ stands for the $M\times M$  identity matrix, and $\mathbf{0}_{M}$ represents a $M\times 1$ vector whose elements are all zero.
\section{System Model and Background}
\label{sysmodel}
\begin{figure}[!h]
    \centering
    \includegraphics[width=8.3cm, height=4.3cm]{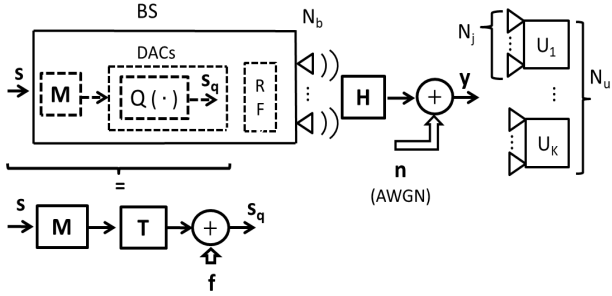}
    \vspace{-1.0em}
    \caption{Outline of a quantized massive MU-MIMO downlink system. Upper diagram: some simplified parts of BS. Lower diagram: Bussgang's theorem applied to the detached part of interest.}
    \label{sysmodel_CQA_BD}
\end{figure}

Let us take into account a BS containing $N_{b}$ antennas, which sends radio frequency (RF) signals to $N_{u}=\sum_{j=1}^{K}\: N_{j}$ receive antennas, where $N_{j}$ denotes the number of receive antennas per $j$th user $U_{j}$, $ j=1,\ldots,K $, as outlined in Fig. \ref{sysmodel_CQA_BD}.
\vskip-1.5ex
We can model the input-output relation of the broadcast channel (BC) as
\begin{equation}
\label{downlink_channel_model}
\mathbf{y}= \mathbf{H}\;\mathbf{s}_{q}\:+\:\mathbf{n},
\end{equation}
where $\mathbf{y} \in \mathbb{C}^{N_{u}} $ contains the signals received by all users and $\mathbf{H} \in \mathbb{C}^{N_{u} \times N_{b}} $ stands for the matrix which models the assumed broadcast channel that is assumed known to the BS. The entries of $\mathbf{H}$ are considered independent circularly-symmetrical complex Gaussian random variables $ \left[ \mathbf{H}\right]_{u,b}\in \mathbb{CN} \left( 0,1\right) $, $u=1,\cdots, N_{u}$ and $b=1,\cdots, N_{b}$. The  noise vector $\mathbf{n} \in\mathbb{C}^{N_{u}} $,  is characterized by its i.i.d.  circularly-symmetric complex Gaussian entries $n_{u}\in \mathbb{CN} \left( 0,\frac{N_{0}}{2}\right)$. We consider that the noise level is known at BS and so is the sampling rate of  DACs at BS and ADCs at user equipments.
Bussgang's theorem \cite{Bussgang,Sven1} allows us to express quantized signals as linear functions of the quantized information  and  distortion expressions, which have no correlation with the signals undergoing quantization. Therefore, the operations performed by the two blocks encompassed by the braces in the upper part of Fig.\ref{sysmodel_CQA_BD} can be transformed  in the expression composed of the operations outlined in the lower part. Thus, the quantization $\mathrm{Q}\left(\cdot \right) $ of a  precoded symbol vector $\mathbf{Ms}$, where $\mathbf{M}\in \mathbb{C}^{N_{b}\times N_{u}}$ is a precoding matrix and $\mathbf{s}\approx \mathit{N}^{\mathbb{C}}\left(\bm{0}_{N_{u}\times1}, \bm{I}_{N_{u}} \right) $ is the symbol vector, can be expressed by the quantized vector
\begin{equation}
\label{bus_theo1}
\mathbf{s}_{q}=\mathrm{Q}\left(\mathbf{Ms}\right)= \mathbf{TMs+f},
\end{equation}
where the distortion term $\mathbf{f}$ and the symbol $\mathbf{s}$ vectors are uncorrelated.  For the general case, $\mathbf{T}\in \mathbb{R}^{N_{b}\times N_{b}}$ is the diagonal
matrix expressed by
\begin{align}
\label{diag_mat_busg}
\mathbf{T}_{n,n}=& \frac{\alpha\gamma}{\sqrt{\pi}}\mathrm{diag}\left(\mathbf{M M}^{H} \right)^{-1/2}.\nonumber\\& \sum_{l=1}^{J-1} \exp\left(-\gamma^{2}\left( l-\frac{J}{2}\right)^{2} \mathrm{diag}\left(\mathbf{M M}^{H} \right)^{-1}  \right)
\end{align}
where  $n=1\ldots N_{b} $, and $\mathit{J}$ and $\gamma$ stand for the number of levels and the step size of the quantizer, respectively. The regularization factor $\alpha \in \mathbb{R}$, which will be defined in Subsection \ref{subsec_approximations},  has the purpose of satisfying the average power limitation
\begin{equation}
\label{power_constraint}
\mathbb {E}[\parallel\mathbf {Ms}\parallel_{2}^{2}]\leq \mathit{P}
\end{equation}
where $\mathit{P}= SNR \:\; \frac{\mathrm{N}_{0}}{2}$.
\subsection{Achievable sum-rates}
\label{subsec_approximations}
In order to compute approximations of achievable sum-rates, in which $N_{b}$ and $N_{u}$ are sufficiently large and that the error resulting from the combination of multiuser interference (MUI) and the distortion from limited resolution of DACs is considered a Gaussian process, we can assume that  \eqref{diag_mat_busg} is the following scalar matrix:
\begin{equation}
\label{approx_diag_mat}
\mathbf{T}_{n,n}= \delta\:\mathbf{I}_{Nb\times Nb}
\end{equation}
where the  entries of $\mathbf{T}_{n,n}$  are given by the Bussgang scalar factor:
\begin{align}
\label{entries_diag_mat}
\delta= \alpha \gamma\sqrt{\frac{N_{b}}{\pi P}}
 \: \sum_{l=1}^{J-1}\exp\left(-\frac{N_{b} \gamma^{2}}{P}\left( 1-\frac{J}{2} \right)^{2} \right)
\end{align}
in which the regularization factor $\alpha$ for enforcing the power constraint \eqref{power_constraint} is obtained by
\begin{align}
\label{normalization_factor}
 \alpha=& \left( 2\mathrm {N_{b}}\gamma^{2} \left(\left( \frac{J-1}{2}\right)^{2} \right.\right.\nonumber\\&\left.\left. -2\sum_{l=1}^{J-1}  \left( 1-\frac{J}{2} \right)  \Xi \left( \sqrt{2N_{b}\gamma^{2}}\left( 1-\frac{J}{2} \right)\right)\right)\right)^{-1/2}
\end{align}
where $\Xi \left( w \right) =\int_{- \infty}^{w} \frac{1}{\sqrt{2 \pi}}\exp^{-v^{2}/2}\:dv$ \cite{Sven1}  is the distributed function of a Gaussian random variable.

It can be proven via Bussgang's theorem that  assuming the system model in Section \ref{sysmodel}  and  the identity in  \eqref{approx_diag_mat}, the   sum rate provided by CQA-BD and CQA-RBD precoders can be approximated by
\begin{align}
\label{achievable_sum_rate_BD_RBD}
\mathit{C}=& \log_{2}\left\lbrace\det\left[ \mathbf{I}_{Nu} + \delta^{2} \frac{\mathit{SNR}}{\mathit{N}_{u}}\mathbf{\left( HM\right) } \mathbf{\left( HM\right) }^{H}\right.\right.\nonumber\\&\left.\left. \left(\left(1-\delta^{2} \right)\frac{\mathit{SNR}}{\mathit{N}_{u}}\mathbf{\left( HM\right) } \mathbf{\left( HM\right) }^{H} +\mathbf{I}_{Nu}              \right)^{-1}                 \right]  \right\rbrace
\end{align}
where the $\mathit{SNR}$ was defined in \eqref{power_constraint} and $\mathbf{M}$ is the precoding matrix \eqref{conj_precod_matrix}, which is defined in Subsection \ref{brief_review_BD_class}.

It must be highlighted that all process of quantization is concentrated in  Bussgang's factor $\delta$ in \eqref{approx_diag_mat} and \eqref{entries_diag_mat}. This is one of the contributions of this work, i.e., the derivation of a closed form expression for estimating achievable sum rates based on a scalar factor which characterizes a  Bussgang's gain scalar matrix \eqref{approx_diag_mat} that approximates the effects of multi-bit quantization. This derivation is provided in the Appendix. The second contribution of this study, which have not been considered in the literature so far, is the application of the obtained closed form expression to evaluate the performance of the achievable sum rates of our proposed CQA-BD and CQA-RBD precoding algorithms under 2,3 and 4-bit quantization.
\subsection{Review of BD precoding algorithms}
\label{brief_review_BD_class}
It is known \cite{Zu,Stankovic,Spencer1} that BD is a low-rank
technique
\cite{xutsa,delamaretsp,kwak,xu&liu,delamareccm,wcccm,delamareelb,jidf,delamarecl,delamaresp,delamaretvt,jioel,rrdoa,delamarespl07,delamare_ccmmswf,jidf_echo,delamaretvt10,delamaretvt2011ST,delamare10,fa10,lei09,ccmavf,lei10,jio_ccm,ccmavf,stap_jio,zhaocheng,zhaocheng2,arh_eusipco,arh_taes,dfjio,rdrab,locsme,okspme,dcg_conf,dcg,dce,drr_conf,dta_conf1,dta_conf2,dta_ls,damdc,song,wljio,barc,jiomber,saalt,locsme,okspme,lrcc}
that employs SVD to design the precoder, which can be performed in
two stages. The precoder computed in the first stage suppresses (BD)
or attemps to obtain a trade-off between MUI and noise (RBD).
Afterwards parallel or near-parallel single user (SU)-MIMO are
calculated. The precoder computed in the second stage parallelizes
the streams intended for the users.  In this way, a precoding matrix
$ \mathbf{M}_{j}$ related to the \textit{j}th user can be expressed
as a product
 \begin{equation}
 \label{precod_as_product}
 \mathbf{M}_{j}=\mathbf{M}_{j}^{c}\mathbf{M}_{j}^{d}
 \end{equation}
 in which $ \mathbf{M}_{j}^{c} \in \mathbb{C}^{N_{b} \times L_{j}}$ and $ \mathbf{M}_{j}^{d} \in \mathbb{C}^{L_{j} \times N_{j}}$. The constant $L_{j}$ depends on which precoding algorithm is chosen (BD or RBD).
We can express the combined channel matrix $ \mathbf{H} $ and the resulting precoding matrix  $ \mathbf{M} $ as follows:
\begin{equation}
\label{comb_ch_matrix}
 \mathbf{H}=\left[\mathbf{H}_{1}^{T} \mathbf{H}_{2}^{T}\cdots\mathbf{H}_{K}^{T}\right]^{T}\:\in \mathbb{C}^{N_{u} \times N_{b}}
\end{equation}
\begin{equation}
\label{conj_precod_matrix}
\mathbf{M}=\left[\mathbf{M}_{1}^{T} \mathbf{M}_{2}^{T}\cdots\mathbf{M}_{K}^{T}\right]^{T}\:\in \mathbb{C}^{N_{b} \times N_{u}}
\end{equation}
where $ \mathbf{H}_{j} \in \mathbb{C}^{N_{j} \times N_{b}} $ is the channel matrix of the $j$th user. The expression $ \mathbf{M}_{j} \in \mathbb{C}^{N_{b} \times N_{j}} $ represents the precoding matrix of the $j$th user.
For BD precoding  algorithm \cite{Stankovic}, the first factor  in \eqref{precod_as_product} is given by
\begin{equation}
\label{precod_mat_first_BD}
\mathbf{M}_{j}^{c\left(BD \right) }=\overline{\mathbf {W}}_{j}^{\left( 0\right) }
\end{equation}
where $ \overline{\mathbf{W}}_{j}^{\left(0 \right) } $ is obtained by the SVD \cite{Zu} of  \eqref{comb_ch_matrix}, in which the matrix channel of $j$th  user was removed, i.e.:

\begin{align}
\label{channel_matrix_user_exclusion}
\overline{\mathbf{H}}_{j}&=\left[\mathbf{H}_{1}^{T}\cdots \mathbf{H}_{j-1}^{T}\mathbf{H}_{j+1}^{T}\cdots\mathbf{H}_{K}^{T}\right]^{T}\:\in \mathbb{C}^{\overline{N}_{j} \times N_{b}}\nonumber\\&= \overline{\mathbf{U}}_{j}\overline{\mathbf{\Phi}}_{j}\overline{\mathbf{W}}_{j}^{H}=\overline{\mathbf{U}}_{j}\overline{\mathbf{\Phi}}_{j}\left[\overline{\mathbf{W}}_{j}^{\left( 1\right) }\overline{\mathbf{W}}_{j}^{\left(0 \right) } \right]^{H}
\end{align}
where $\overline{N}_{j} =N_{u}-N_{j} $.
The matrix $ \overline{\mathbf {W}}_{j}^{\left( 0\right) }
 \in \mathbb{C}^{\overline{N}_{b}
    \times \left( N_{b}-\overline{L}_{j}\right)}$, where $\overline{L}_{j}$ is the assumed rank of $\overline{\mathbf{H}}_{j}$,  embraces the ultimate $ N_{b}-\overline{L}_{j} $ zero singular vectors.
In the case  of RBD precoding algorithm, the first factor in \eqref{precod_as_product} is given \cite{Stankovic} by
\begin{align}
\label{precod_mat_first_RBD}
\mathbf{M}_{j}^{c\left(RBD \right)}= \overline{\mathbf{W}}_{j} \left(\overline{ \mathbf{\Phi}}_{j}^{T} \overline{ \mathbf{\Phi}_{j}} + \alpha \mathbf{I}_{N_{b}}
\right)^{-1/2}
\end{align}
where $ \alpha = \frac{N_{u}\sigma_{n}^{2}}{\mu} $ is the regularization factor and $ \mu $ is the whole average transmit power.

The second factor in \eqref{precod_as_product} is obtained by SVD  of the effective channel matrix for the $j$th user $\mathbf{H}_{e_{j}}$ and power loading, respectively as follows:
\begin{equation}
\label{second_factor_BD}
\mathbf{M}_{j}^{d\left(BD \right) )}=\mathbf{W}_{j}^{\left(1 \right)}\:\Gamma^{\left(BD \right) }
\end{equation}
\begin{equation}
\label{second_factor_RBD}
\mathbf{M}_{j}^{d\left(RBD \right) )}=\mathbf{W}_{j}\:\Gamma^{\left(RBD \right) }
\end{equation}
where the matrix $\mathbf{W}_{j}^{\left(1 \right)}$ embraces the early $ \Lambda_{e}=rank\left( \mathbf{H}_{e_{j}}\right)  $ singular vectors obtained by the decomposition of $ \mathbf{H}_{e_{j}} $, as follows
\begin{align}
\label{effect_channel_matrix}
\mathbf{H}_{e_{j}}&= \mathbf{H}_{j} \mathbf{M}_{j}^{c}\nonumber= \mathbf{U}_{j}\mathbf{\Phi}_{j}\mathbf{W}_{j}^{H}\\&=\mathbf{U}_{j}
\begin{bmatrix}
\mathbf{\Phi}_{j} & 0\\ 0 & 0
\end{bmatrix}
\begin{bmatrix}
\mathbf{W}_{j}^{\left(1 \right)}&\mathbf{W}_{j}^{\left(0 \right)}
\end{bmatrix}^{H}
\end{align}
The power loading matrix $ \Gamma $ can be obtained by a procedure like water filling (WF)\cite{Paulraj}.
\subsection{Increasing requirement of more efficient DACs}
\label{increasing_requirement_efficiency}

Until recently, the use of a modest number of antennas at the BS and their required DACs were not an issue in terms of energy consumption. This is due the fact that DACs consume less energy than ADCs. Despite the diversity of research about DACs, very few allow the calculation of the increment of chip power dissipation as  bit resolution increases bit-by-bit, for a fixed technology. In order to roughly compare the consumption of both equipments, we make use of Table \ref{table1}, which contains the fabrication parameters for GaAs  4-bit  Analog-to-Digital Converter (AD) and  5-bit Digital-to-Analog (DA) converters, using a 0.7-$\mu$m MESFET self-aligned gate process
\cite{Naber} and the expression proposed in \cite{Orhan}.
\begin{table}[htb!]
    \centering
    \caption{ADC and DAC fabrication parameters with the same technology}
    \begin{tabular}{|c c c c |}
        \hline
         & Resolution  & Sampling Rate & Power dissipation \\         [0.5ex]
         & (bits)   & (GHz) &  (mW)  \\
        \hline
        ADC & 4 & 1 & 140 \\
        \hline
        DAC & 5 & 1 & 85 \\ [0.5ex]
        \hline
    \end{tabular}
    \label{table1}
\end{table}

We start with the  expression\cite{Orhan} which relates the power consumed by an ADC  to the resolution bits, as follows:
\begin{equation}
\label{adc_power}
\mathrm{P_{ADC} (b)}= \mathrm{c\:\tau\: 2^{b}}
\end{equation}
where $ b $ stands for the resolution bits, $c$ is a constant and $\tau$ is the sampling rate.
From \eqref{adc_power}, we obtain $\frac{\mathrm{P_{ADC}} (4)}{\mathrm{P_{ADC}} (5)}=\frac{1}{2}$, which allows us to estimate  $\frac{\mathrm{P_{DAC}} (5)}{\mathrm{P_{ADC}} (5)}= \frac{\mathrm{P_{DAC}} (5)}{2\mathrm{P_{ADC}} (4)}$. With the help of Table \eqref{table1}, we obtain $\mathrm{P_{DAC}} (5)\approx 30\% \mathrm{P_{ADC}} (5)$.
So, the DAC consumes around 30 $\% $ of
the energy  of the ADC with fixed parameter. From the results obtained before, we can roughly estimate the economy in energy by assuming that similarly to ADC, DAC consumption doubles with every extra bit of resolution, i.e., of $\mathcal{O}\left(2^{b} \right) $. Therefore, a decrease in two resolution bits, for instance from 4 to 2 bits, represents a consumption 75$\% $ lower. This reduction of DAC consumption motivates our study.
\section{Proposed CQA-BD and CQA-RBD precoder algorithm}
\label{proposed_CQA_algorithms}
In this section, we summarize the proposed precoding techniques in Algorithm \ref{algorithm:CQA_BD_RBD}, which encompasses both CQA-BD and its refined variation CQA-RBD. The algorithm starts with the use of the knowledge of the combined channel matrix \eqref{comb_ch_matrix}. Then, for a fixed SNR and  also a fixed realization of the channel, we perform  the calculations from $1$ to $12$ step-by-step, to obtain the precoding matrix $\mathbf{M} $ \eqref{conj_precod_matrix}. All operations involved in these steps are detailed in Subsection \ref{brief_review_BD_class}, which reviews the BD-type precoding algorithms. Next, we calculate the conformation  parameter $\alpha$ \eqref{normalization_factor}, which ensures the power constraint \eqref{power_constraint}, and after this, the Bussgang's scalar factor \eqref{entries_diag_mat}, which concentrates all process of quantization in the scalar matrix \eqref{approx_diag_mat}. Finally, we can obtain the achievable sum-rates for a fixed SNR and a fixed realization of the channel.
\begin{algorithm}[htb!]
    \scriptsize
    \caption{Algorithm for estimating CQA-BD and CQA-RBD  reachable sums } \label{algorithm:CQA_BD_RBD}
    \begin{algorithmic}[1]
        \Require $\mathbf{H}=\left[\mathbf{H}_{1}^{T} \mathbf{H}_{2}^{T}\cdots\mathbf{H}_{K}^{T}\right]^{T}\:\in \mathbb{C}^{N_{u} \times N_{b}}$ \eqref{comb_ch_matrix}
        \For{$\mathrm{j = 1\;}\colon \: K$}
        \State $\overline{\mathbf{H}}_{j}=\left[\mathbf{H}_{1}^{T}\cdots \mathbf{H}_{j-1}^{T}\mathbf{H}_{j+1}^{T}\cdots\mathbf{H}_{K}^{T}\right]^{T}\:\in \mathbb{C}^{\overline{N}_{j} \times N_{b}}$\eqref{channel_matrix_user_exclusion}
        \State $\overline{\mathbf{H}}_{j}=  \overline{\mathbf{U}}_{j}\overline{\mathbf{\Phi}}_{j}\overline{\mathbf{W}}_{j}^{H}=\overline{\mathbf{U}}_{j}\overline{\mathbf{\Phi}}_{j}\left[\overline{\mathbf{W}}_{j}^{\left( 1\right) }\overline{\mathbf{W}}_{j}^{\left(0 \right) } \right]^{H}$ \eqref{channel_matrix_user_exclusion}
        \State $\mathbf{M}_{j}^{c\left(BD \right) }=\overline{\mathbf {W}}_{j}^{\left( 0\right) }$\eqref{precod_mat_first_BD}
        \State $\mathbf{M}_{j}^{c\left(RBD \right) }=\overline{\mathbf{W}}_{j}\left( \overline{\mathbf{\Phi}}_{j}^{T} \overline{\mathbf{\Phi}}_{j} + \alpha \mathbf{I}_{N_{b}} \right)^{-1/2} $\eqref{precod_mat_first_RBD}
        \State $
        \mathbf{H}_{e_{j}}= \mathbf{H}_{j} \mathbf{M}_{j}^{c}= \mathbf{U}_{j}\mathbf{\Phi}_{j}\mathbf{W}_{j}^{H}=\mathbf{U}_{j} \begin{bmatrix}
        \mathbf{\Phi}_{j} & 0\\ 0 & 0
        \end{bmatrix}
        \begin{bmatrix}
        \mathbf{W}_{j}^{\left(1 \right)}&\mathbf{W}_{j}^{\left(0 \right)}
        \end{bmatrix}^{H}\nonumber $\eqref{effect_channel_matrix}

        \State $\mathbf{\Gamma}^{\left(BD,RBD \right)}  \text{by WF \cite{Paulraj} or similar} $

        \State $
        \mathbf{M}_{j}^{d\left(BD \right) )}=\mathbf{W}_{j}^{\left(1 \right)}\:\Gamma^{\left(BD \right) } $ \eqref{second_factor_BD}
        \State $
        \mathbf{M}_{j}^{d\left(RBD \right) )}=\mathbf{W}_{j}\:\Gamma^{\left(RBD \right) } $ \eqref{second_factor_RBD}
        \State $\mathbf{M}_{j}=\mathbf{M}_{j}^{c}\mathbf{M}_{j}^{d}$\eqref{precod_as_product}

        \EndFor
        \State $ \mathbf{M}=\left[\mathbf{M}_{1}^{T} \mathbf{M}_{2}^{T}\cdots\mathbf{M}_{K}^{T}\right]^{T}\:\in \mathbb{C}^{N_{b} \times N_{u}}  $ \eqref{conj_precod_matrix}

        \State$ \begin{aligned}
        \alpha=& \left( 2\mathrm {N_{b}}\gamma^{2} \left(\left( \frac{J-1}{2}\right)^{2} \right.\right.\nonumber\\&\left.\left. -2\sum_{l=1}^{J-1}  \left( 1-\frac{J}{2} \right)  \Xi \left( \sqrt{2N_{b}\gamma^{2}}\left( 1-\frac{J}{2} \right)\right)\right)\right)^{-1/2}
        \end{aligned} $ \eqref{normalization_factor}

        \State$ \begin{aligned}
        \delta= \alpha \gamma\sqrt{\frac{N_{b}}{\pi P}} \sum_{l=1}^{J-1}\exp\left(-\frac{N_{b}\gamma^{2}}{P}\left( 1-\frac{J}{2} \right)^{2} \right) \text{Bussgang's scalar factor}
        \end{aligned} $ \eqref{entries_diag_mat}

        \State $\begin{aligned}
        \mathit{C}=& \log_{2}\left\lbrace\det\left[ \mathbf{I}_{Nu} + \delta^{2} \frac{\mathit{SNR}}{\mathit{N}_{u}}\mathbf{\left( HM\right) } \mathbf{\left( HM\right) }^{H}\right.\right.\nonumber\\&\left.\left. \left(\left(1-\delta^{2} \right)\frac{\mathit{SNR}}{\mathit{N}_{u}}\mathbf{\left( HM\right) } \mathbf{\left( HM\right) }^{H} +\mathbf{I}_{Nu}  \right)^{-1}               \right]  \right\rbrace
        \end{aligned} $ \eqref{achievable_sum_rate_BD_RBD}
    \end{algorithmic}
\end{algorithm}
\section{Numerical results}
\label{numerical_results}
We focus our simulations on two scenarios, which are composed of
$\left( N_{b1}, N_{u1}\right)=\left( 32,16\right)$ and $\left(
N_{b2}, N_{u2}\right)=\left( 128,16\right)$, respectively. We model
the channel matrix $\mathbf{H}_{j}$ of the $j$th user with entries
given by complex Gaussian random variables with zero mean and unit
variance. In addition, it is assumed that the channel is static
while each packet is transmitted and that the antennas are
uncorrelated. The channel estimation is considered ideal at the
receive side and there is no error in the feedback of the channel
information to the receiver. We set the trials to $2\times 10^2$ and
the packet length to $ 10^2$ symbols.

\begin{figure}[htb!]
    \centering
    \includegraphics[width=8.6cm,height=6.4cm]{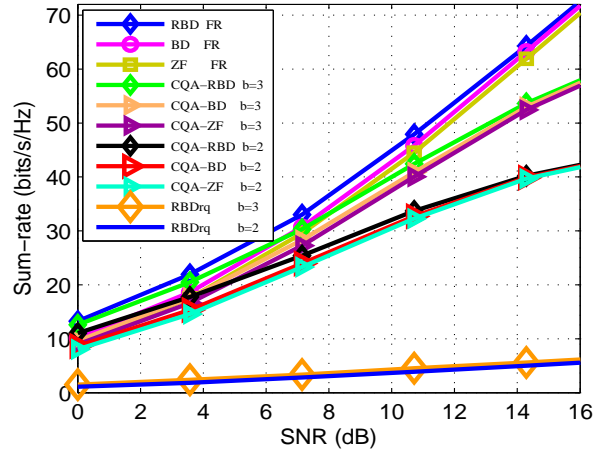}
    \caption{Approximations of reachable rates via Bussgang theorem assuming Gaussian signals for CQA-RBD, CQA-BD and CQA-ZF algorithms, $ N_{b}=32 $  and $ N_{u}=16 $. Decreasing levels of quantization from  the upper  side to the middle  $\left( N_{b1}, N_{u1}\right)=\left( 32,16\right)$. At the bottom,  the roughly quantized RBDrq for $ \mathrm{J}=\left\lbrace  8,4\right\rbrace  $ }
    \label{Sum_18_11_2_3_bit_32_16_composed_3}
\end{figure}

Fig.\ref{Sum_18_11_2_3_bit_32_16_composed_3} depicts the sum-rates of the proposed CQA-BD and CQA-RBD algorithms, based on Bussgang's theorem, under 2 and 3-bit quantization, corresponding to $ \mathrm{J}=\left\lbrace  8,4\right\rbrace$ levels of quantization, and employing the first scenario $\left( N_{b1}, N_{u1}\right)=\left( 32,16\right)$. For the purpose of comparisons, we have also included the sum-rate of CQA-ZF, which represents the same technique used in CQA-BD type, applied to Zero-Forcing (ZF). They are compared to the  RBD (refined variant of BD), the standard BD, and ZF, at upper side,  all of them in full resolution (RBD FR, BD FR, ZF FR). They are also compared to RBD under 2 and 3-bit roughly  standard quantization, i.e., not using Bussgang's theorem. It is possible to notice four well-defined ranked groups of curves in the range $\left[ 7 \:20\right]$ and that,  in the same range, the rank of each group is also  clear. Thus, the inequalities Sum RBD$>$ Sum BD $>$ Sum ZF is preserved, regardless of quantization. It is also clear the influence of the levels of quantization on the  the sum-rates. We highlight the clear gaps between each group of sum-rates achieved corresponding to each level of quantization. It can be noticed the significant increasing gaps from 2-3bit roughly quantized (RBDqr), at the bottom, to 2-3bits and CQA-RBD, and how close 3-bit CQA-RBD is to full resolution (FR) RBD. Based on Subsection \ref{increasing_requirement_efficiency}, the performance of CQA-RBD and CQA-BD under 2-3 bits quantization, which increasingly approximates RBD FR and BD FR, indicating a corresponding saving in energy.

\begin{figure}[htb!]
    \centering 
    \includegraphics[width=8.6cm,height=6.4cm]{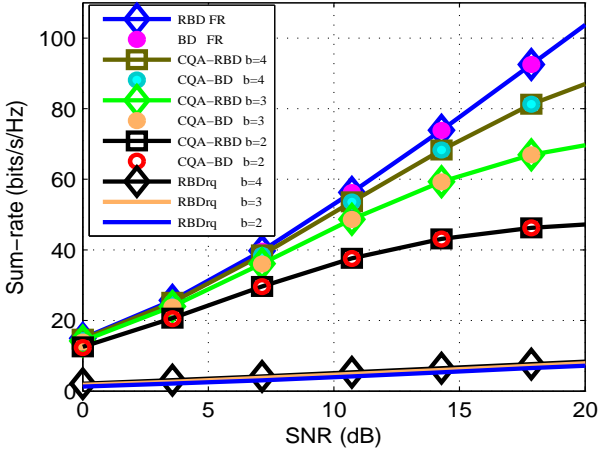}
    \caption{Approximations of reachable rates via Bussgang theorem assuming Gaussian signals for CQA-BD and  CQA-RBD algorithms, $ N_{b}=128 $  and $ N_{u}=16 $. Decreasing levels of quantization from upper side to the middle  $ \mathrm{J} =\left\lbrace \infty, 16,8,4\right\rbrace $. At the bottom,  the roughly quantized RBDrq for $ \mathrm{J}=\left\lbrace  16,8,4\right\rbrace  $ }
    \label{Sum_18_11_2_3_4bit_128_16_composed}
\end{figure}

Fig. \ref{Sum_18_11_2_3_4bit_128_16_composed} illustrates the
performance of the sum-rates in the second configuration mentioned
before, i.e., $\left( N_{b2}, N_{u2}\right)=\left( 128,16\right)$.
In this arrangement, we compare only 2,3,4-bit CQA-BD class to
2,3-bit RBDrq (roughly quantized) and RBD FR,BD FR. It can be
noticed that the curves depicting CQA-RBD and CQA-BD at each level
of quantization are almost similar, which can be justified by the
increase of transmit antennas. However the aim of the figure is to
show more clearly how 2,3,4-bit CQA-RBD and CQA-BD   algorithms
increase sum-rates, comparing them to a bad condition, i.e., 2,3-bit
RBDrq and to a ideal condition RBD-FR. Taking RBD FR as a reference
at 3.57 and 7.14 dB, the sum-rate achieved by  2-bit CQA-RBD,
represents $80\%$ and $74\%$, respectively, of the rates achieved
with a full-resolution system. For 3-bit CQA-RBD, at the same range,
the sum-rate achieves $93\%$ and $90\%$, respectively, of that
achieved by full resolution. This means substantially less energy
dissipated at the cost of slightly lower sum-rates, which justify to
the investigation of low-resolution precoding techniques using 2,3
bits as an alternative of 1-bit quantization-based precoders. Future
work will include the development of detection and decoding
techniques
\cite{mmimo,wence,Li2011,deLamare2003,itic,deLamare2008,cai2009,jiomimo,dfcc,deLamare2013,did,rrmser,jidf,XWang1999,mfsic,mbdf,bfidd,1bitidd,aaidd,dopeg,memd,vfap}

\section{Conclusions}
\label{conclusions}
Founded on Bussgang's theorem, which allows us to express quantized signal as linear functions of the quantized information and distortion, we have proposed the CQA-BD and CQA-RBD precoding algorithms for multi-bit DACs, in particular, using 2,3,4-bit quantization.These approaches have not been taken into account in the literature so far. We have also justified the need for reducing the energy consumption in DACs by using few bits of quantization, as a way of compensating the increase of power dissipation resulting from modern large-scale MIMO systems. Compared to the current full resolution RBD and BED algorithms and coarsely quantized RBD algorithm, the sum-rates obtained by CQA-RBD and BD achieve significant gains.
\appendix
Here we provide the derivation of \eqref{achievable_sum_rate_BD_RBD}.
\subsection{Assumptions}
According to Subsection \ref{sysmodel}, we  assume that the cross covariance matrices $\mathbf{R}_{sf}=\mathbb E[\mathbf{s}\mathbf{f}^{H}]$ and $\mathbf{R}_{sn}=\mathbb E[\mathbf{s}\mathbf{n}^{H}]$ and $\mathbf{R}_{fn}=\mathbb E[\mathbf{f}\mathbf{n}^{H}]$ are all of them equal to zero and so are $\mathbf{R}_{fs}$, $\mathbf{R}_{ns}$ and $\mathbf{R}_{nf}$.
Additionally, as stated in Subsection \ref{approx_diag_mat}, the distortion vector is assumed to be Gaussian.
 \subsection{Development}
 We start by combining \eqref{downlink_channel_model} with \eqref{bus_theo1}, obtaining:
\begin{equation}
\label{downlink_channel_model_mod_Paulraj}
\mathbf{y}= \mathbf{HTMs}+\mathbf{Hf}\:+\:\mathbf{n},
\end{equation}
where we can define a distortion-plus-noise vector
\begin{equation}
\label{distortion_plus_noise_vector}
\mathbf{\tilde{n}}= \mathbf{Hf}\:+\:\mathbf{n}.
\end{equation}
We can estimate the correlation matrix $\mathbf{R}_{sqsq}$ of the quantized vector \eqref{bus_theo1}, as follows:
\begin{align}
\label{autocorrel_quantized_vector}
\mathbf{R}_{sqsq}&=\mathbb E[\mathbf{(TMs+f)}\mathbf{(TMs+f)}^H]\nonumber\\&=  \mathbb E[\mathbf{TMs}\mathbf{s}^{H}\mathbf{M}^{H}\mathbf{T}^{H}+\mathbf{f}\mathbf{f}^{H}]\nonumber\\&=\delta^{2}\sigma^{2} \mathbf{M}\mathbf{M}^{H} +\mathbf{R}_{ff},
\end{align}
where we made use of \eqref{approx_diag_mat} and the autocorrelation matrix of the symbol vector  $\mathbf{R}_{ss}=\mathbb E[\mathbf{s}\mathbf{s}^{H}]=\sigma_{s}^{2}\mathbf{I}_{Nu}$, in which  $ \sigma_{s}^{2} $ is its variance. The term  $\mathbf{R}_{ff}=\mathbb E[\mathbf{f}\mathbf{f}^{H}]$ stands for the autocorrelation of the distortion vector $\mathbf{f}$.

Next, we can notice that in full resolution, since there is no quantization and its associated distortion, \eqref{bus_theo1} turns into
\begin{equation}
\label{bus_theo1_reduced}
\ \mathbf{s}_{q}=\mathbf{Ms},
\end{equation}
where we make $\mathbf{T}_{n,n}= \mathbf{I}_{Nb\times Nb}$, i.e., $\delta= 1$  in \eqref{approx_diag_mat}, and assume that $\mathbf{f}=\textbf{0}_{Nb}$.

Now, we calculate the autocorrelation of the full resolution precoded symbol vector \eqref{bus_theo1_reduced} as follows:
\begin{align}
\label{autocorrel_fr_precod_symbol_vector}
\mathbf{R}_{sqsq}&=\mathbb E[\mathbf{(Ms)}\mathbf{(Ms)}^H]\nonumber\\&=  \mathbb E[\mathbf{Ms}\mathbf{s}^{H}\mathbf{M}^{H}]\nonumber\\&=\sigma_{s}^{2} \mathbf{M}\mathbf{M}^{H},
\end{align}
By equating \eqref{autocorrel_quantized_vector} and \eqref{autocorrel_fr_precod_symbol_vector}, we can obtain the expression of the autocorrelation of the distortion vector $\mathbf{R}_{ff}$:
\begin{align}
\label{autocorr_distort}
&\delta^{2}\sigma_{s}^{2} \mathbf{M}\mathbf{M}^{H} +\mathbf{R}_{ff}=\sigma_{s}^{2} \mathbf{M}\mathbf{M}^{H}\nonumber\\&\therefore \mathbf{R}_{ff}=\left(1- \delta^{2}\right)\sigma_{s}^{2}\mathbf{M}\mathbf{M}^{H}
\end{align}
By equating \eqref{autocorrel_quantized_vector} and \eqref{autocorrel_fr_precod_symbol_vector}, we can obtain the expression of the autocorrelation of the distortion vector $\mathbf{R}_{ff}$:
\begin{align}
\label{autocorr_distort}
&\delta^{2}\sigma_{s}^{2} \mathbf{M}\mathbf{M}^{H} +\mathbf{R}_{ff}=\sigma_{s}^{2} \mathbf{M}\mathbf{M}^{H}\nonumber\\&\therefore \mathbf{R}_{ff}=\left(1- \delta^{2}\right)\sigma_{s}^{2}\mathbf{M}\mathbf{M}^{H}
\end{align}
We can then compute the autocorrelation matrix of \eqref{downlink_channel_model_mod_Paulraj}, obtaining:
\begin{align}
\label{cov_matrix_channel_model_mod_Paulraj}
\mathbf{R}_{yy}&=\mathbb E[\mathbf{y}\mathbf{y}^H] \nonumber\\&= \left( \mathbf{HTM}\right) \:\mathbf{R}_{ss}\left( \mathbf{HTM}\right) ^{H}\nonumber\\&+\mathbf{H}\:\mathbf{R}_{ff}\mathbf{H}^{H}\:+\:\mathbf{R}_{nn},
\end{align}
where $\mathbf{R}_{ss}=\mathbb E[\mathbf{s}\mathbf{s}^{H}]$,  $\mathbf{R}_{ff}=\mathbb E[\mathbf{f}\mathbf{f}^{H}]$ and $\mathbf{R}_{nn}=\mathbb E[\mathbf{n}\mathbf{n}^{H}]$ are the autocorrelation matrices of the signal,  the distortion and the noise vectors, respectively.
Similar procedure applied to the distortion-plus-noise vector \eqref{distortion_plus_noise_vector}, considering the conditions above, yields:
\begin{align}
\label{cov_matrix_dist_plus_noise}
\mathbf{R}_{\tilde{n}\tilde{n}}=\mathbb E[{\tilde{\mathbf{n}}\tilde{\mathbf{n}}^H}]=\mathbf{H}\mathbf{R}_{ff}\mathbf{H}_{H} +\mathbf{R}_{nn},
\end{align}
From the principles of information theory \cite{Cover} and the capacity of a  frequency flat deterministic MIMO channel \cite{Paulraj}, we can bound  the achievable rate in bits per channel use at which information can be sent with arbitrarily low probability of error by the mutual information of a Gaussian channel, i.e.
\begin{align}
\label{reachable_rate}
\mathit{C} \leqq \mathit{I}\left(\mathbf{s},\mathbf{y} \right)&= \Upsilon\left(\mathbf{y} \right)-\Upsilon\left(\mathbf{y}\rvert\mathbf{s} \right)\nonumber\\&=\Upsilon\left(\mathbf{y} \right)-\Upsilon\left(\mathbf{\tilde{n}}\right)\nonumber\\&= \log_{2}\left[ \det\left( \pi e\mathbf{R}_{yy}\right) \right] -\log_{2}\left[ \det\left( \pi e\mathbf{R}_{\tilde{n}\tilde{n}}\right) \right]\nonumber\\&= \log_{2}\left[ \det\left( \mathbf{R}_{yy}\right) \right] -\log_{2}\left[ \det\left(  \mathbf{R}_{\tilde{n}\tilde{n}}\right) \right]\nonumber\\&=
\log_{2}\left[ \det\left( \mathbf{R}_{yy}\mathbf{R}^{-1}_{\tilde{n}\tilde{n}}\right) \right]
\end{align}
where $\Upsilon\left(\mathbf{y}\right) $  and $\Upsilon\left(\mathbf{y}\rvert\mathbf{s}_{q} \right)$ are the differential and the conditional differential entropies of  $\mathbf{y}$, respectively.

By combining \eqref{cov_matrix_channel_model_mod_Paulraj} and \eqref{cov_matrix_dist_plus_noise} with \eqref{reachable_rate}, we have:
\begin{align}
\label{reachable_rate_2}
\mathit{C} &\leqq\log_{2}\left\lbrace \det\left[ \left( \left( \mathbf{HTM}\right) \:\mathbf{R}_{ss}\left( \mathbf{HTM}\right) ^{H}\right)\right.\right.\nonumber\\&\left.\left.\left(\mathbf{H}\mathbf{R}_{ff}\mathbf{H}^{H} +\mathbf{R}_{nn} \right)^{-1} +\mathbf{I}_{N_{u}}
\right] \right\rbrace
\end{align}
In Section \ref{sysmodel}, \eqref{power_constraint}, we have defined the total power as  $\mathit{P}= \mathit{SNR}\: \frac{\mathrm{N}_{0}}{2}$. From the definition of the noise vector, also in that Section, we can express  its covariance matrix as $\mathbf{R}_{nn}=\frac{\mathrm{N}_{0}}{2}\mathbf{I}_{N_{u}} $
Combining the two previously mentioned expressions, we obtain
\begin{align}
\label{noise_covar_matrix}
\mathbf{R}_{nn}=\frac{\mathit{P}}{SNR}\;\mathbf{I}_{N_{u}}
\end{align}
Recalling, from Section \ref{sysmodel}, that $\mathbf{R}_{ss}\approx \mathbf{I}_{N_{u}} $, and  assuming that the total power  is given by $\mathit{P}= trace\left( \mathbf{R}_{ss}\right)= N_{u}  $, \eqref{noise_covar_matrix} turns into:
\begin{align}
\label{noise_covar_matrix_mod}
\mathbf{R}_{nn}=\frac{\mathit{N_{u}}}{SNR}\;\mathbf{I}_{N_{u}}
\end{align}
By combining \eqref{reachable_rate_2}, \eqref{autocorr_distort} \eqref{approx_diag_mat} and the expression of $\mathbf{R}_{ss}$ previously mentioned with \eqref{noise_covar_matrix_mod}, followed by algebraic manipulation, we can obtain:
\begin{align}
\label{achievable_sum_rate_BD_RBD_final}
\mathit{C}=& \log_{2}\left\lbrace\det\left[ \mathbf{I}_{Nu} + \delta^{2} \frac{\mathit{SNR}}{\mathit{N}_{u}}\mathbf{\left( HM\right) } \mathbf{\left( HM\right) }^{H}\right.\right.\nonumber\\&\left.\left. \left(\left(1-\delta^{2} \right)\frac{\mathit{SNR}}{\mathit{N}_{u}}\mathbf{\left( HM\right) } \mathbf{\left( HM\right) }^{H} +\mathbf{I}_{Nu}              \right)^{-1}                 \right]  \right\rbrace \qed
\end{align}

\end{document}